\documentclass{PoS}
\usepackage{latexsym}
\usepackage{amssymb}
\usepackage{epsf}
\usepackage{amsmath}
\usepackage{color}
\usepackage{graphicx}
\usepackage{epstopdf}

\usepackage{slashed}

\newcommand{\figcaption}[1]{\def\@captype{figure}\caption{#1}}
\newcommand{\tblcaption}[1]{\def\@captype{table}\caption{#1}}
\newcommand{\Nf}{N_f}
\newcommand{\kl}{\kappa_l}
\newcommand{\kh}{\kappa_h}

\newcommand{\ml}{m_l}

\def\simge{\mathrel{%
       \rlap{\raise 0.511ex \hbox{$>$}}{\lower 0.511ex \hbox{$\sim$}}}}
\def\simle{\mathrel{
       \rlap{\raise 0.511ex \hbox{$<$}}{\lower 0.511ex \hbox{$\sim$}}}}

\title{Many flavor approach to study the nature of chiral phase
transition of two-flavor QCD}

\ShortTitle{Many flavor approach to study the nature of chiral phase
transition of two-flavor QCD}

\author{
\speaker{Norikazu Yamada}\\
KEK Theory Center, Institute of Particle and Nuclear Studies, High
Energy Accelerator Research Organization (KEK), Tsukuba 305-0801,
Japan\\
School of High Energy Accelerator Science, SOKENDAI (The Graduate
University for Advanced Studies), Tsukuba 305-0801, Japan\\
E-mail: \email{norikazu.yamada@kek.jp}
}

\author{
Shinji Ejiri\\
Department of Physics, Niigata University, Niigata 950-2181, Japan\\
E-mail: \email{ejiri@muse.sc.niigata-u.ac.jp}
}

\author{
Ryo~Iwami\\
Graduate School of Science and Technology, Niigata University, Niigata
950-2181, Japan\\
E-mail: \email{iwami@muse.sc.niigata-u.ac.jp}
}

\abstract{
We perform lattice numerical simulations to study the phase transition
of QCD at finite temperature to clarify the nature of the transition of
massless two flavor QCD.
We investigate QCD with two light and $\Nf$ heavy quarks instead of
two-flavor QCD, and focus on the light quark mass dependence of the
critical heavy mass, below which the transition is of first order.
The heavy quarks are incorporated into two flavor configurations in the
form of the hopping parameter expansion through the reweighting
technique.
The nature of the transition is identified by the shape of the
constraint effective potential at the critical temperature.
Our result indicates that the critical heavy mass remains finite in
the chiral limit of the two flavors, suggesting the phase transition of
massless two-flavor QCD is of second order.
}

\FullConference{The 33rd International Symposium on Lattice Field Theory\\
		14 -18 July 2015\\
		Kobe International Conference Center, Kobe, Japan*}

\begin{document}

\section{Introduction}
 \label{sec:introduction}

Since the outstanding work by Pisarski and
Wilczek~\cite{Pisarski:1983ms}, clarifying the nature of chiral phase
transition of massless two flavor QCD has been one of the central
problems in QCD.
The nature crucially depends on the presence of the flavor singlet axial
symmetry $U_A(1)$ at $T_c$~\cite{Pisarski:1983ms}, and the nature itself
and the effective restoration of $U_A(1)$ above $T_c$ are under active
investigation~\cite{Bonati:2009yg,Aoki:2012yj,Cossu:2013uua,Pelissetto:2013hqa,Nakayama:2014sba,Bonati:2014kpa,Sato:2014axa}.
Clarifying this point is important not only for understanding the phase
structure of QCD but also for the axion dark matter scenario.
The axion abundance is essentially determined by the temperature
dependence of the topological susceptibility $\chi_t$, which vanishes
when $U_A(1)$ symmetry is effectively and fully restored.
If $\chi_t$ vanishes very rapidly right above $T_c$, too much axions
would be produced, and the axion dark matter scenario becomes hard or is
even excluded, depending on how rapidly it
happens~\cite{Kitano:2015fla}.
The lattice studies of the axion dark matter scenario recently
began in the quenched
approximation~\cite{Kitano:2015fla,Berkowitz:2015aua,Borsanyi:2015cka}.

\begin{figure}[h]
\begin{center}
\begin{tabular}{cc}
\includegraphics*[width=0.45 \textwidth,clip=true]
{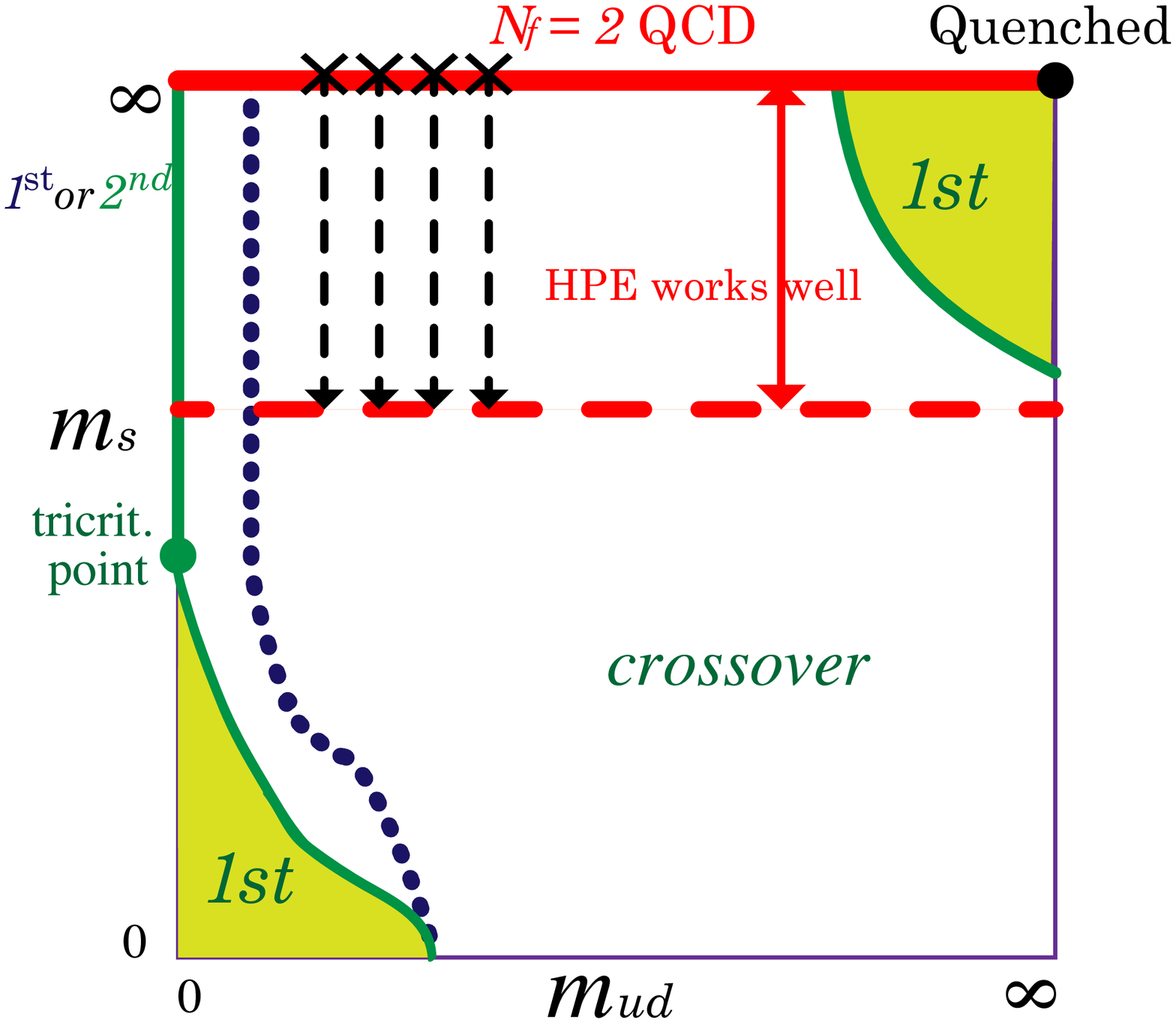}&
\includegraphics*[width=0.45 \textwidth,clip=true]
{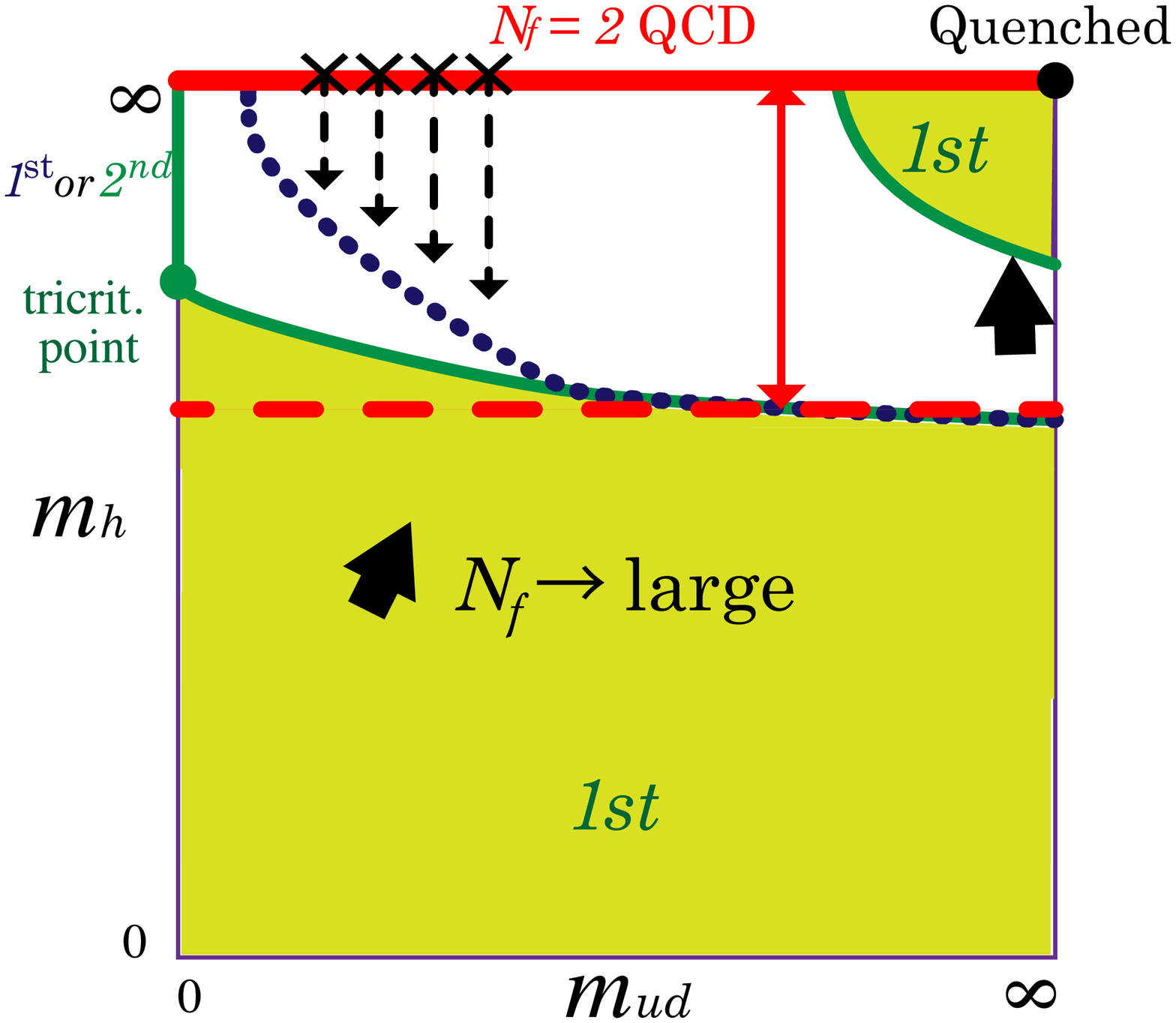}\\
 (a)\ 2+1 flavor QCD&
 (b)\ 2+$\Nf$ flavor QCD\\
\end{tabular}
\caption{Basic idea of many flavor approach.}
\vspace{-2ex}
\label{fig:columbia-plot}
\end{center}
\end{figure}
In this work, we apply the approach proposed in
Ref.~\cite{Ejiri:2012rr}.
The essential idea is depicted in Fig.~\ref{fig:columbia-plot}.
Figure~\ref{fig:columbia-plot} (a) shows the so-called Columbia plot for
2+1 flavor QCD. 
There are two distinct first order regions lying around the quenched
limit ($m_{ud}=m_s=\infty$) and the chiral limit of three flavor QCD
($m_{ud}=m_s=0$), respectively.
In the following, we focus only on the latter.
The question is whether the massless two flavor QCD point ($m_{ud}=0$
and $m_s=\infty$) is inside the first order region or not.
If we could trace the critical line (solid or dotted curve), the
question is resolved.
However, it is difficult as the critical line is located in the small
quark mass region.

The situation becomes tractable by adding extra flavors.
Figure~\ref{fig:columbia-plot} (b) represents the expected Columbia plot
for 2+$\Nf$ flavor QCD, where $m_h$ denotes the mass of extra $\Nf$
flavors.
In this case, the critical line moves upward, and for sufficiently
large $\Nf$ the critical line becomes reachable by the hopping parameter
expansion (HPE)~\cite{Ejiri:2012rr}.
If the critical heavy mass $m_h^c(m_{ud})$ turns out to remain finite in
the $m_{ud}\to 0$ limit, it immediately means that massless two flavor
QCD corresponding to the point $(m_{ud},m_h)=(0,\infty)$ is the outside
of the first order region.
An important remark is that, in the limit of $m_s\to\infty$ or
$m_h\to\infty$, both of 2+1 and 2+$\Nf$ flavor QCD end up with the same
two flavor QCD.
Thus, the original question is simplified to whether the critical heavy
mass in the chiral limit of two flavors stays finite or not.
In the following, we briefly describe our approach and the result.
For details, pleasee see the full paper~\cite{Ejiri:2015vip}.

\section{method}
\label{sec:method}

Two flavor configurations at finite temperatures are generated following
the standard Hybrid Monte Carlo method at four values of two flavor
mass.
The effective potential $V$ is obtained from the probability
distribution function (PDF) $w$ for the generalized plaquette
$P$~\cite{Ejiri:2007ga,Ejiri:2008xt,whot11,Ejiri:2012rr} as
\begin{eqnarray}
       V(P;\beta,\kl,\kh,\Nf)
&=&  - \ln w(P;\beta,\kl,\kh,\Nf)
 \label{eq:effective-potential}\\
    w(P; \beta, \kl, \kh,\Nf)
&=& \int \!\! {\cal D} U\, \delta(P- \hat{P}) \ 
    \big[\det M(\kh)\big]^{\Nf}
    e^{-S_{\rm gauge}(\beta) - S_{\rm light}(\kl)},
\label{eq:pdist}\\
    \hat P
&=& c_0\, \hat W_P + 2 c_1\, \hat W_R\ ,
\end{eqnarray}
where
$S_{\rm gauge}(\beta)= 6\, N_{\rm site}\,\beta\,
 \big\{ (c_0+2c_1) - \hat P\big\}$
and $S_{\rm light}(\kl)$ are the lattice actions for the gauge field and
two flavors of light quarks, respectively.
$\hat W_P$ and $\hat W_R$ denote the averaged plaquette and rectangle,
respectively, and $c_0$ and $c_1$ satisfying $c_0=1-8 c_1$ are the
improvement coefficients for lattice gauge action.
$M(\kappa_h)$ is the quark matrix for heavy flavors.
When measuring the PDF, $\Nf$ flavors of extra heavy quarks are
introduced in the form of the hopping parameter expansion (HPE) via the
reweighting method as shown later.

We separate the effective potential into two parts for convenience as
\begin{eqnarray}
    V(P; \beta_{\rm ref},\kl,\kh,\Nf)
&=&   V_{\rm light}(P; \beta_{\rm ref},\kl)
    - \ln R(P;\beta_{\rm ref},\kl,\kh,\Nf)\ .
\label{eq:vefftrans}
\end{eqnarray}
Here and hereafter, terms independent of $P$ are ignored, because we are
interested in the differentiation of the potential with regard to $P$.
The first term is defined by
\begin{eqnarray}
    V_{\rm light}(P; \beta_{\rm ref},\kl)
= - \ln w(P;\beta,\kl,0,0)  
  - 6\,N_{\rm site}\,(\beta_{\rm ref} - \beta)\,P
\label{eq:V-two-flavor}
\end{eqnarray}
and represents the constraint effective potential in two flavor QCD
system.
The second term of eq.~(\ref{eq:vefftrans}) is defined by
\begin{eqnarray}
    R(P;\beta_{\rm ref},\kl,\kh,\Nf)
&=& \left\langle
    \displaystyle
    \left[ \det M(\kh) \right]^{\Nf}
    \right\rangle_{P: {\rm fixed},(\beta_{\rm ref},\kl)}, \ 
\label{eq:lnr}\\
    \langle \cdots \rangle_{P: {\rm fixed}, (\beta_{\rm ref},\kl)}
&\equiv&
    \frac{\langle \delta(P- \hat{P}) \cdots \rangle_{(\beta_{\rm ref},\kl)}}
         {\langle \delta(P- \hat{P}) \rangle_{(\beta_{\rm ref},\kl)}}\ ,
\end{eqnarray}
where $\langle \cdots \rangle_{(\beta,\kl)}$ denotes the ensemble
average over two flavor configurations generated with $\beta$ and $\kl$.
Then, assuming the unimproved Wilson Dirac operator for $M(\kh)$ with a
sufficiently small $\kh$ and $N_t=4$, $\ln R$ in
eq.~(\ref{eq:vefftrans}) can be approximated as
\begin{eqnarray}
    \ln R(P;\kl,h)
&\approx&
    \ln \left\langle
    \displaystyle
    \exp\left( 6\,N_s^3\,h\,\hat Y \right)
    \right\rangle_{P: {\rm fixed},(\beta,\kl)}
\label{eq:r-1}\\
&=&  9\,N_{\rm site}\,\frac{h}{c_0}\, P
   + \ln R'(P;\kl,h)
\label{eq:r-2}
\end{eqnarray}
where
\begin{eqnarray}
    \ln R'(P;\kl,h)
&=& \ln \left\langle
    \displaystyle
    \exp\left( 6\,N_s^3\,h\,\hat Z \right)
    \right\rangle_{P: {\rm fixed},(\beta,\kl)}\ .
\label{eq:rdash}
\end{eqnarray}
In the above, the following quantities have been introduced,
\begin{eqnarray}
    h
&=& 2\,\Nf\,(2 \kh)^4\ ,\ \ \
    \hat Y
= 6\,\hat W_P + \hat{L}\ ,\ \ \
    \hat Z
= - \frac{12\,c_1}{c_0}\,\hat W_R + \hat{L}\ ,
\label{eq:h}
\end{eqnarray}
with $L$ the real part of Polyakov loop averaged over the spatial
volume.
Although eqs.~(\ref{eq:r-1}) and (\ref{eq:r-2}) are algebraically
identical, the equality is not necessarily trivial in numerical data
because the $\delta$ function is approximated by
\begin{eqnarray}
        \delta(x)
 \approx 1/(\Delta \sqrt{\pi}) \exp[-(x/\Delta)^2]\ .
 \label{eq:delta-func}
\end{eqnarray}
Thus, both expressions are examined to check the consistency.
At this order of the HPE, the number of extra heavy flavors ($N_f$)
and their mass parameter ($\kh$) appear only in a single parameter $h$.
The nature of the phase transition is identified by the shape of the
constraint effective potential, {\it i.e.} single- or double-well, at
$T=T_c$.
By scanning $h$, we determine the critical value $h_c$, at which the
first order and the crossover regions are separated.

With our definition of the PDF, one can prove that the curvature of the
effective potential is independent of
$\beta_{\rm ref}$~\cite{Ejiri:2007ga,Ejiri:2012rr}, which greatly
simplifies the analysis (For detail, see Ref.~\cite{Ejiri:2015vip}).
By looking at the light quark mass dependence of $h_c(\kl)$, we try to
extract $h_c$ in the chiral limit.

\section{results}
\label{sec:numerical_results}

Following Ref.~\cite{whot10}, we take the Iwasaki gauge action
($c_1=-0.331$) and the $O(a)$-improved Wilson fermion action with the
perturbatively improved $c_{\rm sw}$ for two flavors of light quarks.
Simulations are performed on $N_{\rm site}=16^3\times 4$ lattices with
25 to 32 $\beta$ values at each $\kl$,
and 10,000 to 40,000 trajectories have been accumulated at each
simulation point.
Four light quark masses are ranging from $\kl=0.145$ to 0.1505,
corresponding to being from $0.46 < m_\pi/m_\rho < 0.66$.

\begin{figure}[tb]
\begin{center}
\begin{tabular}{cc}
\includegraphics*[width=0.5 \textwidth,bb=5 5 360 240,clip=true]
{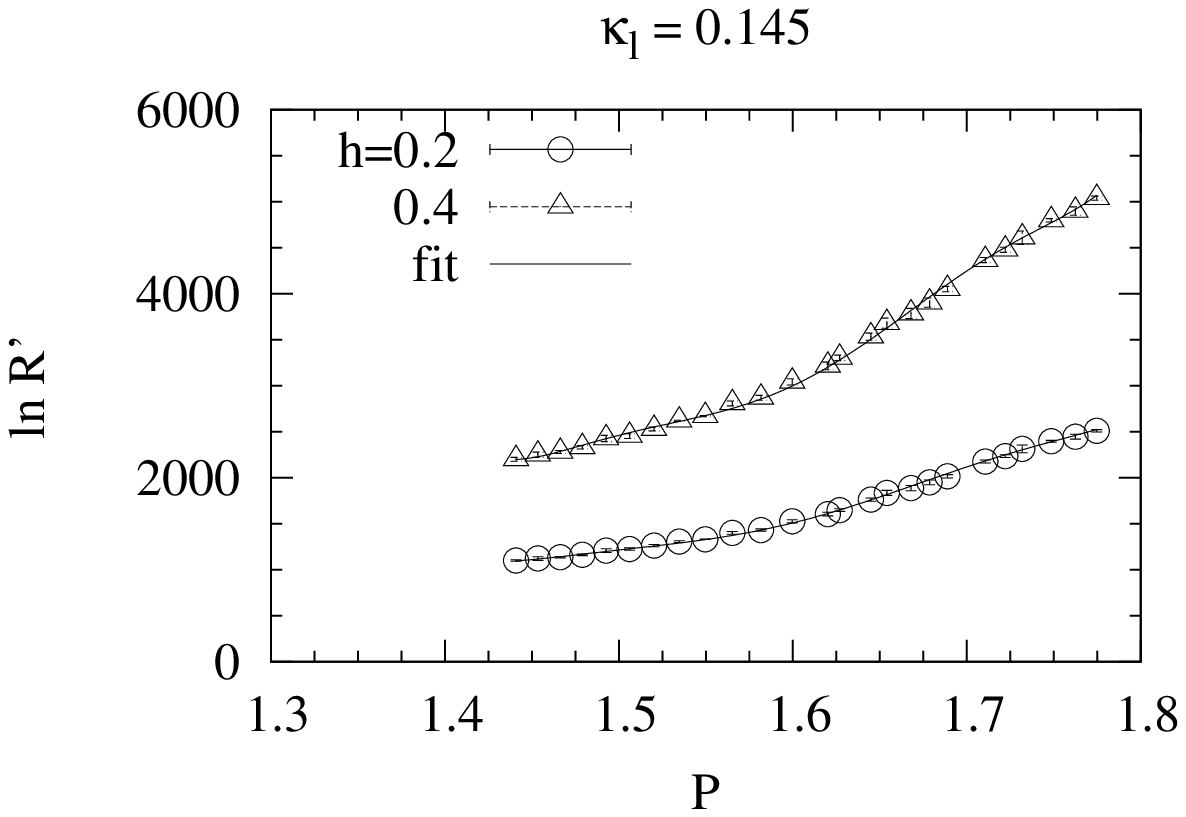}&
\includegraphics*[width=0.5 \textwidth,bb=5 5 360 240,clip=true]
{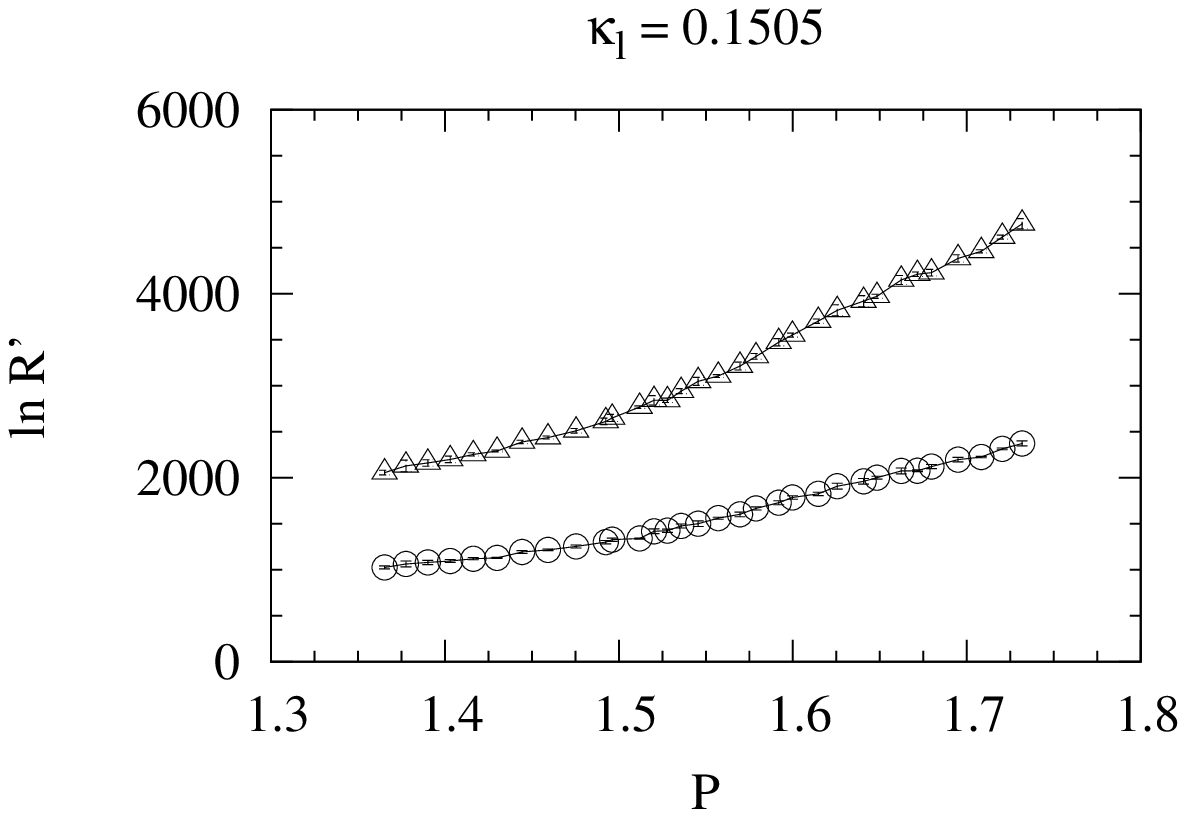}\\
\end{tabular}
\vspace{-3ex}
\caption{\
 $\ln R'$
 at $\kl=0.145$  (left) and 0.1505 (right).
 The results at $h$=0.2 to 0.4 are shown.
 }
\label{fig:lnr-Xsp2}
\vspace{-2ex}
\end{center}
\end{figure}

In the approximation of $\delta$, we take two values of $\Delta$ =
0.0001 and 0.00025 to see the stability of the results, and the
discrepancy arising from different choice is taken as a systematic
uncertainty.
In the following plots, the results with $\Delta$=0.0001 are shown
unless otherwise stated.
As an example, the $P$ dependence of $\ln R'(P;\kl,h)$ are shown
in Fig.~\ref{fig:lnr-Xsp2}.
The statistical errors are invisible in this scale.

We fit the data of $\ln R$ and $\ln R'$ to polynomial functions of $P$.
The fits are made over three fit ranges, three different polynomial
orders,  two values of $\Delta$.
Since not all the fits are successful, we only keep the fit results
satisfying $\chi^2/{\rm dof} < 3$ in the following analysis.
Furthermore, it turns out that five or six parameters are enough to
fit the data well, thus we omit the fits with the sixth order
polynomial.
Once the fit parameters are determined, it is straightforward to
calculate the curvature of the potential.

\begin{figure}[h]
\begin{center}
\begin{tabular}{cc}
\includegraphics*[width=0.5 \textwidth,bb=0 0 360 240,clip=true]
{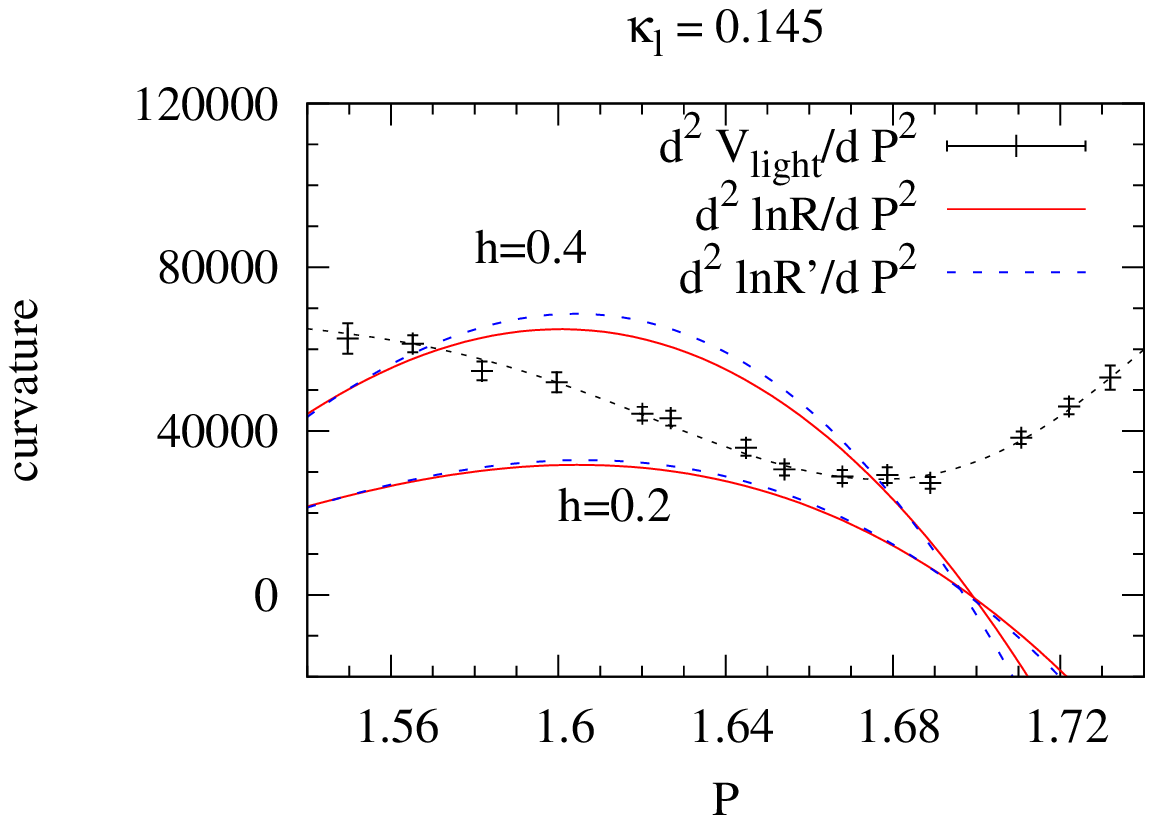}&
\includegraphics*[width=0.5 \textwidth,bb=0 0 360 240,clip=true]
{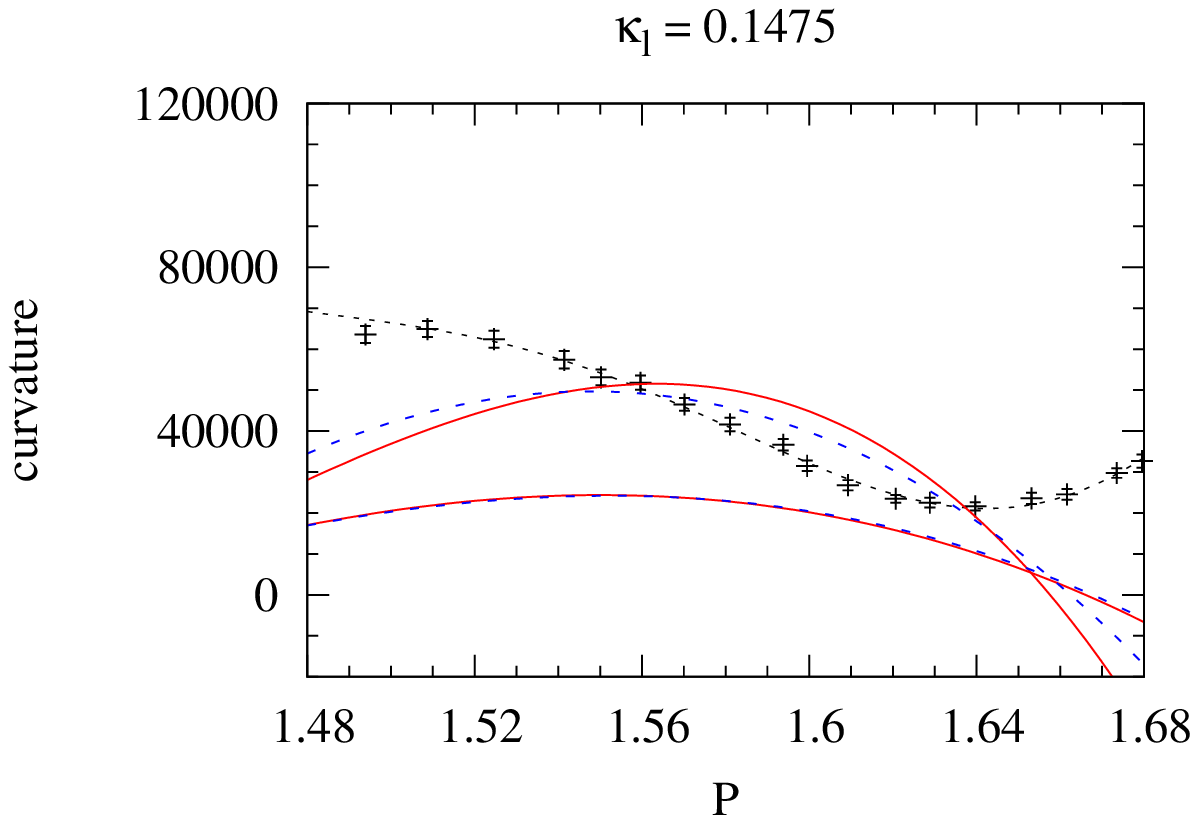}\\
\includegraphics*[width=0.5 \textwidth,bb=0 0 360 240,clip=true]
{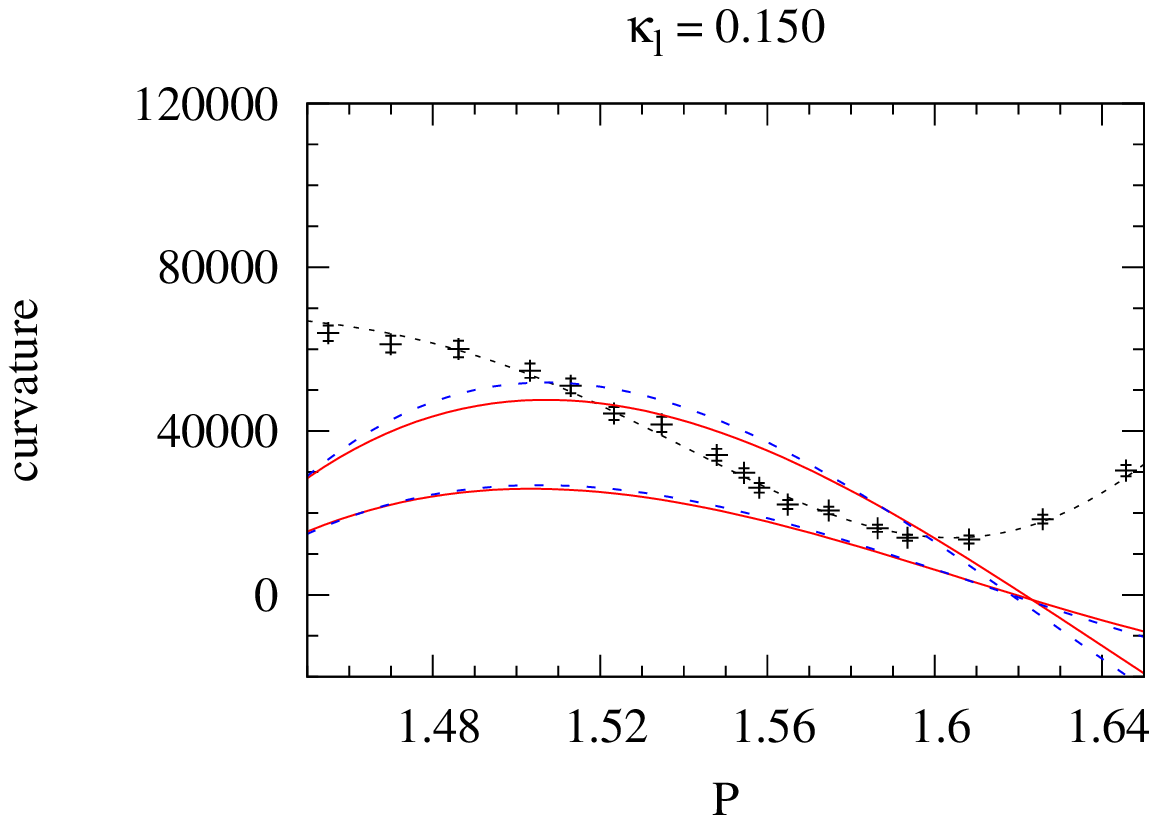}&
\includegraphics*[width=0.5 \textwidth,bb=0 0 360 240,clip=true]
{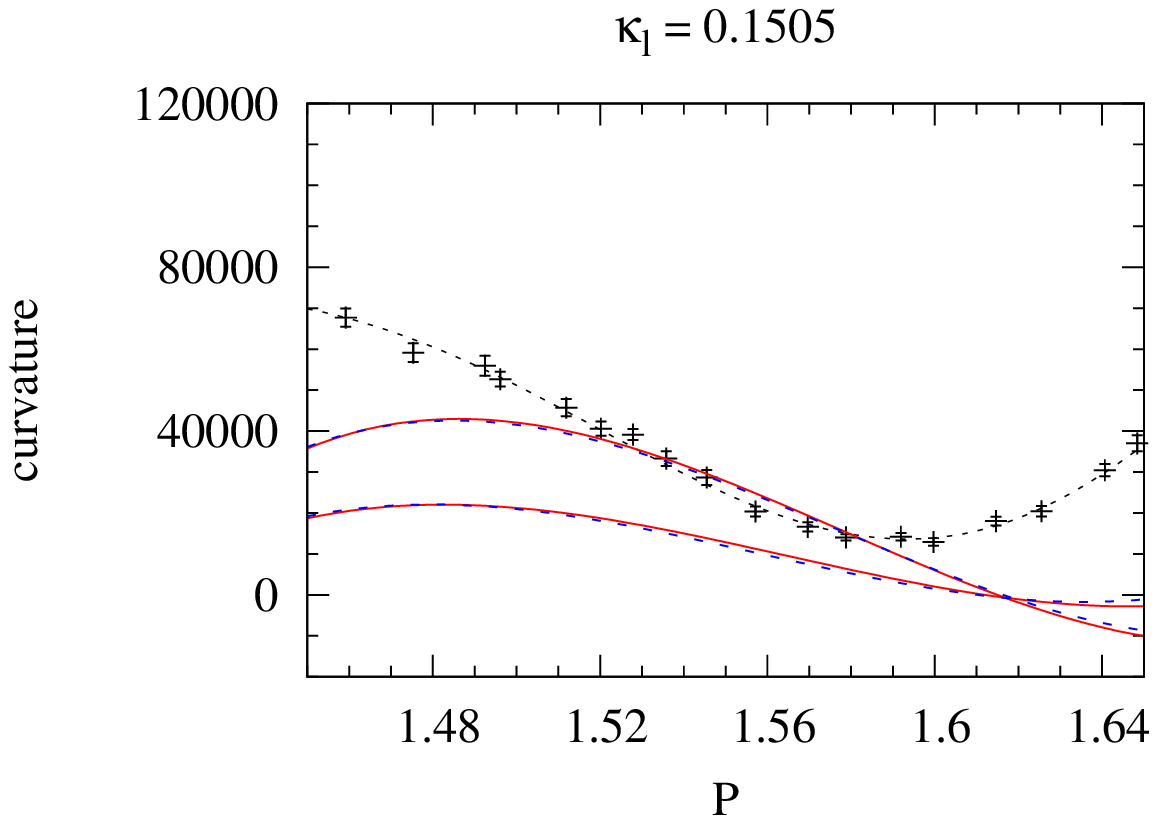}\\
\end{tabular}
\vspace{-2ex}
\caption{
 The second derivative of the first term and the second term of the
 effective potential are shown as a function of $P$.
 The second term contribution exceeds that of the first term in a range
 of $P$ when $h=0.4$, which indicates the occurrence of the first order
 transition at such a value of $h$.
 }
\vspace{-2ex}
\label{fig:d2lnr}
\end{center}
\end{figure}
The results for the curvature are plotted in Figs.~\ref{fig:d2lnr} for
$\ln R$ (solid curves) and $\ln R'$ (dashed curves), where the results
for $h=0.2$ and 0.4 are shown as examples.
The difference in the curvature between $\ln R$ and $\ln R'$ turns out
to be reasonably small at all $\kl$.

The curvature of the first term of eq.~(\ref{eq:V-two-flavor}) can be
calculated using the averaged value and the susceptibility of $\hat P$
at each $\beta$~\cite{Ejiri:2015vip}, which is shown in
Fig.~\ref{fig:d2lnr} with the dotted curves obtained from the fit to a
fifth order of polynomial.
It is seen that, independently of $\kl$, $d^2V_{\rm light}/dP^2$ is
always positive as expected.
The figure shows that $d^2 \ln R/dP^2$ and $d^2 \ln R'/dP^2$ have a
peak at slightly below the $P$ value at which $d^2V_{\rm light}/dP^2$
takes the minimum, which indicates that, in many flavor system, the
phase transition or rapid crossover occurs at $P$ smaller than the
two flavor case.

\begin{figure}[bt]
\begin{center}
\begin{tabular}{c}
\includegraphics*[width=0.6 \textwidth,bb=0 0 360 240,clip=true]
{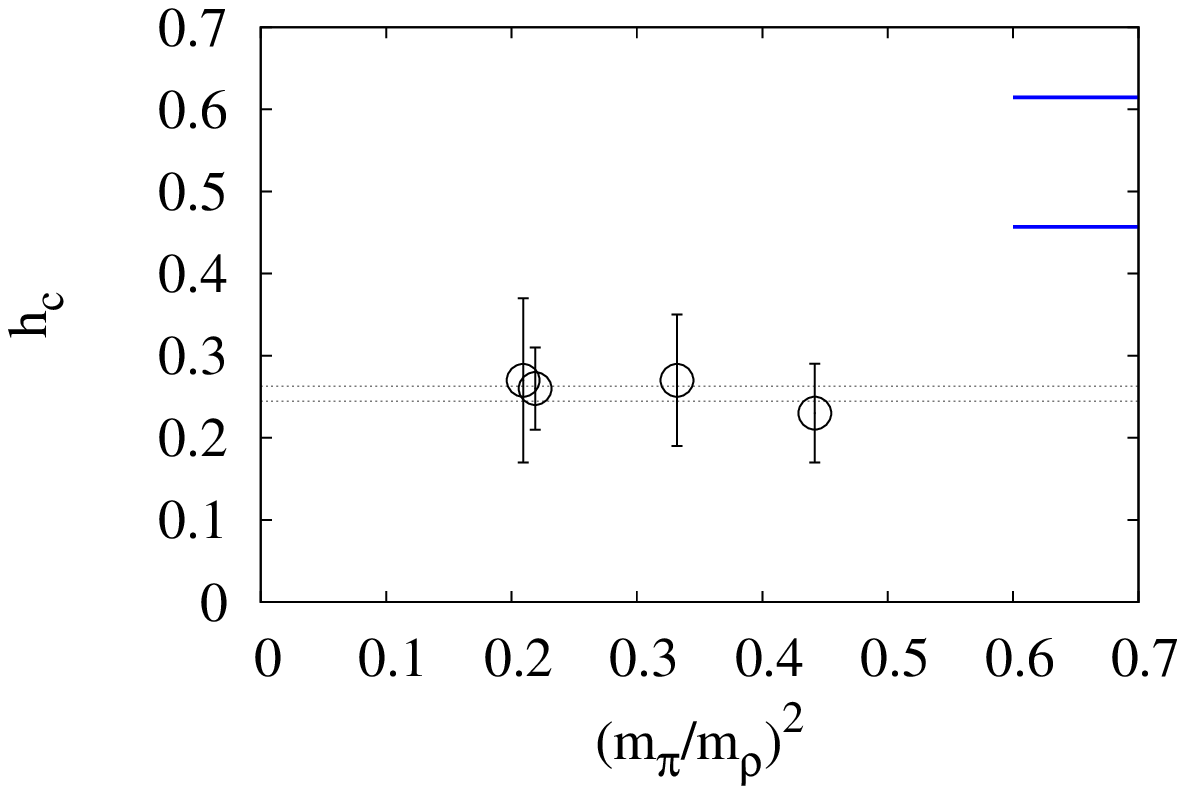}
\end{tabular}
\vspace{-2ex}
\caption{The light quark mass dependence of $h_c$.
 The solid lines represent a constant fit, where the band on the top
 right corner is $h_c$ obtained from the $2+\Nf$ flavor simulation with
 $\kl=0.0$ and $\Nf=50$.
 }
\vspace{-2ex}
\label{fig:lq-dep}
\end{center}
\end{figure}

To determine $h_c$, we iterated the calculation with $h$ varying in
steps of 0.02.
It turns out that the resulting $h_c$ depends on the details
of the fitting procedures, though not by much.
We adopt all those results as long as $\chi^2/dof<3$, and the
spread is taken as the systematic uncertainty.
Figure~\ref{fig:lq-dep} shows the light quark mass dependence of $h_c$,
where the error bars represent the systematic uncertainties.
Since no clear tendency is observed in the light quark mass dependence,
we fit the data to a constant (solid lines), yielding $h_c=0.23(1)$ in
the chiral limit of the two flavor mass.
Nonzero value of $h_c$ in the chiral limit excludes the possibility of
the first order transition of massless two flavor QCD.

\section{Summary and discussion}
\label{sec:summary}

We have studied the finite temperature phase transition of QCD with two
light and many heavy quarks at zero chemical potential to identify the
critical line separating the continuous crossover and the first order
regions on the $\kl$-$h$ plane.
In other words, two flavor QCD with a finite mass is enforced to undergo
a first order transition by adding extra quarks.
We then try to see whether those extra quarks are necessary to keep the
first order transition down to the chiral limit of two flavor mass.

The nature of the transition is identified by the shape of the
constraint effective potential for the generalized plaquette.
It is that $h_c$ stays constant within the systematic uncertainty in
the range of two flavor mass we have studied
($0.46 \le m_\pi/m_\rho\le 0.66$), which suggests that the critical
heavy mass remains finite in the chiral limit of the two flavors and
hence the phase transition of massless two flavor QCD is of second
order.

Our approach is applicable for any kinds of light quark action.
$\kh$ can be considered to be arbitrary small by assuming arbitrary
large $\Nf$.
Therefore, the above statement is valid independently of the convergence
of the hopping parameter expansion.

To establish our finding, it is crucial to confirm the tricritical
scaling~\cite{Ukawa:1995tc,ejirilat08,Ejiri:2013lia},
\begin{eqnarray}
h_c \sim (\mbox{const.})\times \ml^{2/5} + \mbox{const.} ,
 \label{eq:meanfield-scaling}
\end{eqnarray}
where the power $2/5$ is independent of $\Nf$.
In order to confirm this power, it is essential to reduce the systematic
uncertainties associated with the fitting procedure.

We can extend this approach to explore QCD at finite chemical potential
as initiated in Ref.~\cite{Ejiri:2012rr,Iwami}.
We believe that such a study brings useful information in understanding
rich QCD phase structure.

\section{Acknowledgments}

We would like to thank members of WHOT-QCD Collaboration for
discussions.
We also thank Ken-Ichi Ishikawa
for providing us his simulation codes.
This work is in part supported by
JSPS KAKENHI Grant-inAid for Scientific Research (B)
(No.\ 15H03669 [NY],
      23540295 [SE]
)
and (C)
(No.\ 26400244 [SE]), and by the Large Scale Simulation Program of High
Energy Accelerator Research Organization (KEK) No.\ 14/15-23.

\end{document}